\documentclass[draft,linenumbers]{agujournal}
\usepackage{url} 
\usepackage{enumitem}
\usepackage{amsmath}
\usepackage{multirow}
\usepackage{makecell}
\usepackage{setspace}

\def\thickline{\noalign{\vskip 3pt}\Xhline{3\arrayrulewidth}\noalign{\vskip 3pt}}

\draftfalse

\journalname{Geophysical Research Letters}

\begin{document}

\title{Testing a proposed ``second continent'' beneath eastern China using geoneutrino measurements}

\authors{Bed\v{r}ich Roskovec \affil{1}, Ond\v rej \v Sr\'amek \affil{2}, William F. McDonough \affil{3,4,5}}

\affiliation{1}{Instituto de F\'{i}sica, Pontificia Universidad Cat\'{o}lica de Chile, Santiago, Chile}
\affiliation{2}{Department of Geophysics, Faculty of Mathematics and Physics, Charles University, Prague, Czech Republic}
\affiliation{3}{Research Center for Neutrino Science, Tohoku University, 6-3, Aramaki Aza Aoba, Aobaku, Sendai, Miyagi, 980-8578, JAPAN}
\affiliation{4}{Department of Geology, University of Maryland, College Park, MD 20742, USA}
\affiliation{5}{Department of Earth Sciences, Tohoku University, 6-3, Aramaki Aza Aoba, Aobaku, Sendai, Miyagi, 980-8578, JAPAN}

\vspace{10 pt}

\correspondingauthor{Bed\v{r}ich Roskovec}{beroskovec@uc.cl}

\begin{abstract}
Models that envisage successful subduction channel transport of upper crustal materials below 300 km depth, past a critical phase transition in buoyant crustal lithologies, are capable of accumulating and assembling these materials into so-called ``second continents'' that are gravitationally stabilized at the base of the Transition Zone, at some 600 to 700~km depth. Global scale, Pacific-type subduction (ocean-ocean and ocean-continent convergence), which lead to super continent assembly, were hypothesized to produce second continents that scale to about the size of Australia, with continental upper crustal concentration levels of radiogenic power. Seismological techniques are incapable of imaging these second continents because of their negligible difference in seismic wave velocities with the surrounding mantle. We can image the geoneutrino flux linked to the radioactive decays in these second continents with land and/or ocean-based detectors. We present predictions of the geoneutrino flux of second continents, assuming different scaled models and we discuss the potential of current and future neutrino experiments to discover or constrain second continents. The power emissions from second continents were proposed to be drivers of super continental cycles. Thus, testing models for the existence of second continents will place constraints on mantle and plate dynamics when using land and ocean-based geoneutrino detectors deployed at strategic locations.
\end{abstract}

\section{Introduction}
Subduction of tectonic plates leads to the recycling of eroded continental materials back into the mantle, where entrained, down-going crustal materials are transported beyond the mantle magmatic zone that is the source of arc magmas. In recent years, many have observed \citep[e.g.,][]{HUENE,SCHOLL,YAMAMOTO2009443,STERN2011284} that deep crustal recycling has added more than a crustal mass of continental material back into the mantle. Deeply subducted crustal material may be concentrated at either the bottom of the mantle Transition Zone or at the core-mantle boundary, with the buoyancy of this crustal material depending on its bulk compositional characteristics, which in turn dictates it mineralogy and hence density. Concentration of more granitic like lithologies are envisaged as being density stabilized at the bottom of the mantle Transition Zone at some 600 to 700~km depth \citep{MARU2011,KAWAI2013, SAF_MARU}, forming so-called second continents (SC). Being derived from the crust, these SC are enriched in heat-producing elements (i.e., long-lived radionuclides $^{40}$K, $^{232}$Th, and $^{238}$U). Global scale, Pacific-type subduction are presumed to produce SC that can be about the size of Australia, with continental upper crustal concentration levels of radiogenic power \citep{KAWAI2013}. Moreover, it has also been suggested that such a strongly localized heat source can reduce the time scale of continental drift \citep{ICHIKAWA2013}. It has also been proposed that SC can also supply and trigger volatile-bearing plumes \citep{SAFONOVA}. Over time, due to thermal heating, the viscosity of these SC masses will cause these accreted domains to flow and disperse, potentially producing uniform layer at the base of the Transition Zone.

In principle, the SC could be observed by seismic methods. However, trade-offs between composition and temperature, and the uncertainty of these parameters preclude a clear interpretation of the discrepancy between seismic observations and seismic velocity data from laboratory experiments on mantle minerals' held at condtions equivalent to the bottom of the transition zone \citep{KAWAI2013}. Along with the heat production, SC will emit vast numbers of mostly electron antineutrinos, called geoneutrinos, from the radioactive decays of K, Th and U \citep{KRAUSS}. The SC would be a bright source of geoneutrinos due to the high content of these radionuclides. Geoneutrinos can be detected by large antineutrino detectors that are in operation today or proposed to be built.

The leading method of detection of electron antineutrinos is inverse beta decay (IBD): $\bar{\nu}_e+p\rightarrow e^++n$ with a threshold of $E_{\bar\nu_e}^{thr}=1.8$~MeV. This energy restriction allows us to detect only antineutrinos produced during some of the $\beta^-$ decays in the $^{238}$U and $^{232}$Th decay chains, since only these decays have energies that exceeds the IBD threshold. Geoneutrino fluxes have been measured by the KamLAND \citep{KAMLAND} and BOREXINO \citep{BOREXINO} experiments. New detectors will be coming online in the near future, in particular SNO+ \citep{SNO}, JUNO \citep{JUNO}, and Jinping \citep{JINPING}. In addition to these land-based experiments, there is a proposal to build a mobile ocean bottom detector HanoHano \citep{OBK}, which can provide complementary measurements at various ocean bottom locations.

We propose that geoneutrino measurements can successfully test the presence of a SC. A strong SC geoneutrino signal would represent an excess signal on top of the predicted flux that does not take into account a contribution from a SC. In principle, the power of a single experimental measurement is limited, given the uncertainties. However, the scrutinizing power can be significantly improved by comparing signals from closely located experiments or, in the case of an ocean bottom movable detector, by comparing measurements at different locations.

This letter is organized as follows: We describe the prediction of a geoneutrino flux at a given location assuming no SC. We then set up and calculate model predictions for SC with expected location, size of the continent, and abundances of radionuclides. Following that, we calculate the expected flux for current and future geoneutrino experiments and highlight the predicted additional signal due to presence of a SC. We then discuss the potential of discovering a second continent or, assuming no deviation from classical prediction is observed, placing limits on its size and position. Finally, we present conclusions and recommendations regarding other potential applications.

\section{Geoneutrino Flux Prediction}
A geoneutrino flux prediction is based on a global Earth model that describes the Earth's geometry, density and abundance of radionuclides. We adopt the modeling approach used by \citep{SRAMEK} and use the same abundances and distribution of radionuclides in the crust and the mantle. The CRUST1.0 model \citep{CRUST} is used to describe the crust's physical structure. Each $1^\circ(\text{lat})\times1^\circ(\text{lon})$ crustal column is assigned a tectonothermal province of either oceanic or continental type and is vertically defined by two water layers (if present) and six rocky layers, with each layer having an assigned thickness, density and average seismic speeds. Below the crust is the continental lithospheric mantle that extends down to a common depth of 175~km. The underlying mantle is assumed to be spherically symmetrical with two layers:  upper depleted and lower enriched mantle. The mantle density is taken from the PREM model \citep{PREM}. Each layer type is given an average $^{238}$U and $^{232}$Th abundance. The mean values and uncertainties are listed in Tab.~\ref{tab:abun}. The global model assumes uranium and thorium abundances to be correlated within a layer. Abundances in lithosphere are treated as uncorrelated among layers, however, correlation of lithosphere and convective mantle layers is introduced by constraining the total heat production of the Earth to be $(20\pm4)$~TW \citep{MCDONOUGH1995}. Such a compositional model results in the so-called Medium-Q convecting mantle model with $\sim$13~TW radiogenic heat power, which is mildly favoured by geoneutrino flux measurements. Other Earth compositional estimates lead to Low-Q and High-Q mantle models with $\sim$3~TW and $\sim$23~TW respectively. We focus on the Medium-Q model in this paper, however, we take into account all models in evaluation of the uncertainty for the classical prediction and assessment of the SC discovery potential.

\begin{table}[ht]
\begin{center}
\caption{Mean values and uncertainties of element abundances for each layer type.  \label{tab:abun}}
    \scriptsize{
    \begin{tabular}{ c  c  c  c}  \thickline
    Layer Type			&			$A_{Th}$ (kg/kg)		&			$A_{U}$ (kg/kg) & Reference	\\ \thickline
    \begin{tabular}{c}Sediments and\\ Upper Continental Crust \end{tabular}&	$(10.5\pm1.05)\times10^{-6}$					&		$(2.7\pm0.57)\times10^{-6}$		& \citet{RUDNICK20031}		\\ \hline
    Middle Continental Crust	&			$(6.5\pm0.52)\times10^{-6}$			&			$(1.3\pm403)\times10^{-6}$	&	\citet{RUDNICK20031}	\\ \hline
    Lower Continental Crust	&			$(1.2\pm0.36)\times10^{-6}$			&		$(0.2\pm0.06)\times10^{-6}$	&	\citet{RUDNICK20031}		\\ \hline
    Oceanic Sediments &			$(8.1\pm0.57)\times10^{-6}$			&		$(1.73\pm0.087)\times10^{-6}$		&	\citet{plank:2014tgc}	\\ \hline
    Oceanic Crust &				$(210\pm63)\times10^{-9}$		&		$(70\pm21)\times10^{-9}$		&	\citet{white:2014tgc}	\\ \hline
    \begin{tabular}{c}Continental \\ Lithospheric Mantle \end{tabular} &		$150^{+277}_{-97})\times10^{-9}$				&	$33^{+49}_{-20}\times10^{-9}$			&		\citet{Huang}\\ \hline
    Depleted Mantle &					$(21.9\pm4.38)\times10^{-9}$	&			$(8.0\pm1.6)\times10^{-9}$		&	\citet{AREVALO201070}\\ \hline
    Enriched Mantle* &					$147^{+74}_{-57}\times10^{-9}$	&		$30^{+24}_{-18}\times10^{-9}$		&		 Calculated from mass balance \\  \thickline
    \end{tabular}
    }
    *Enriched mantle is assumed to be 19\% of total convective mantle mass \citep{arevalo:2013}, with abundances calculated to balance the amount of radionuclides in the Earth model of \citet{MCDONOUGH1995}.
\end{center}
\end{table}

Given the global Earth model, we calculate the expected geoneutrino differential flux $\frac{d\phi(\vec{r},E)}{dE}$ as a function of antineutrino energy at the detector location $\vec{r}$ as:
\begin{equation}
\frac{d\phi(\vec{r},E)}{dE}=\sum_i{\frac{X_i N_A}{\tau_i\mu_i}\left(\frac{dN_{\bar\nu}}{dE}\right)_i\langle P_{osc} \rangle \int{\frac{A_i(\vec{r}')\rho(\vec{r}')}{4\pi\vert \vec{r}-\vec{r}'\vert}d\vec{r}'},}
\label{eq:fluxcal}
\end{equation}
where the integration goes over the Earth volume with $\vec{r}'$ specifying the position of the antineutrino emitter and the sum includes the detectable radionuclides. We use the averaged neutrino oscillation probability $\langle P_{osc} \rangle$ neglecting the small effects of the exact formula at distances $\leq$60~km. Variables and their values are explained in Tab.~\ref{tab:variables}.

\begin{table}[ht]
\begin{center}
\caption{\label{tab:variables}Description and values of variables used in geoneutrino flux calculation in Eq.~\ref{eq:fluxcal} for two relevant radionuclides.}
\begin{tabular}{cccc} \thickline
Symbol & Description & $^{232}$Th & $^{238}$U \\ \thickline
$X$	&	Natural isotopic mole fraction	&	1.0	&	0.993	\\ \hline
$\mu$	&	Standard atomic weight [g/mol]	&	232.04	&	238.03	\\ \hline
$\tau$ [y]	&	Mean lifetime	&	$2.03\times10^{10}$	&	$6.45\times10^9$	\\ \hline
$N_A[mol^{-1}]$	&	Avogadro constant	&	\multicolumn{2}{ c }{$6.022\times10^{23}$}	\\ \hline
$\langle P_{osc} \rangle$	&	Average oscillation probability	&	\multicolumn{2}{ c }{0.533}	\\ \hline
$\frac{dN_{\bar\nu}}{dE}$	&	$\bar\nu_e$ energy spectrum	&	\multicolumn{2}{ c }{From \citet{ENOMOTO}}	\\ \hline
$A(\vec{r})$	&	Element abundance	&	\multicolumn{2}{ c }{ Based on Tab.~\ref{tab:abun}}	\\ \hline
$\rho(\vec{r})$	&	Matter density	&	\multicolumn{2}{ c }{CRUST1.0 and PREM}	\\ \thickline
\end{tabular}
\end{center}
\end{table}

The geoneutrino flux is often expressed in terrestrial neutrino units (TNU), which corresponds to the number of geoneutrinos detected via IBD on $10^{32}$ free protons (equivalent to a 1 kiloton detector) for a one year long exposure with 100\% detection efficiency. The flux in TNU is then expressed as:
\begin{equation}
\Phi(\vec{r})[\text{TNU}]=3.09\times10^{39}\int{\frac{d\phi(\vec{r},E)}{dE}\sigma_{\text{IBD}}(E)dE},
\label{eq:phi_tnu}
\end{equation}
where $\sigma_{\text{IBD}}(E)$ is the IBD cross-section \citep{VOGEL}. Integration in Eq.~\ref{eq:phi_tnu} effectively goes from IBD threshold to about 3.3~MeV, where geoneutrino spectrum ends \citep{ENOMOTO}.

\section{Second Continent Scenarios}
The size of a SC and its abundances of radionuclides can vary, based on the presumed scenario. A newly created SC might be found at the top of the lower mantle and bottom of the Transition Zone (i.e., 600 to 700 km depth). The SC will be situated in front of the down-going slab that projects back to the subduction zone trench where oceanic plates are thrust beneath continental plates. The size of SC will scale with the width of the down-going plate, the subduction channel efficiency, and the lifetime of the subduction zone. Following \citet{KAWAI2013} and \citet{SAF_MARU}, we assume the abundances of uranium and thorium are comparable to that in the upper continental crust. Second continents older than $\sim$1~Ga would disperse over large areas but be maintained at basal Transition Zone depths, in general no longer linked to the current position of the subduction zones. Given sufficient time, the SC could evolve as a viscously spreading gravity current into a uniform layer around the Earth. Element abundances in these ancient SC would be, in principle, lower than in the newly created SC due to diffusion and mixing with ambient Transition Zone materials.

In order to investigate the possibility to detect a SC with geoneutrinos, we have tested several SC scenarios for which we calculated the predicted geoneutrino flux and discus the implications for measurements done with the current and future neutrino detectors. We consider these models:
\begin{itemize}
\item \textbf{Model~Ia:} A homogeneous layer around the Earth at 600--700~km depth. We assume Upper Continental Crust abundances $A_{\text{Th}}=10.5\times10^{-6}\text{ kg/kg}$ and  $A_{\text{U}}=2.7\times10^{-6}\text{ kg/kg}$, see Tab.~\ref{tab:abun}, without any constraint on total mass of U and Th in the Earth.
\item \textbf{Model Ib:} We constrain total mass of radionuclides in the Earth to the results in \citet{MCDONOUGH1995}. While having crustal model from \v{S}r\'{a}mek et al. (2016), we assume that all uranium and thorium present in the convective mantle is localized in the SC layer with geometry from Model~Ia, the rest of the mantle being void of Th and U. This is unlike the classical scenario where we assume layers of depleted and enriched mantle.
\item \textbf{Model~II:} Following the models presented in  \citet{MARU2011,KAWAI2013,SAF_MARU,SAFONOVA}, we assume an Australia-size SC. The SC lateral shape is initially defined as a ``square'' in latitude?-longitude on a curvilinear grid, centered at position 0º(lat) 0º(lon), with dimensions of 2000~km by 2000~km measured along great circle paths and 100 km thick placed at 600~km depth. Subsequently it was moved to its desired location under China, centered at $32^\circ$N $111^\circ$E (see left panel of Fig.~\ref{fig:sc_asia}) and placed at a depth of 600--700~km. Such a SC could be a result of the Pacific plate subduction under the Asian continent. The abundances are set to $A_{\text{Th}}=10.5\times10^{-6}\text{ kg/kg}$ and  $A_{\text{U}}=2.7\times10^{-6}\text{ kg/kg}$, with no constraints on the total radionuclide mass in the Earth. The additional mass of uranium and thorium is about 5\% of the total mass in the Earth.
\item \textbf{Model~III:} We assume an Australia-size SC with the same shape of $2000\times2000$~km as in Model~II located in the South Atlantic, centered at $44^\circ$S $47^\circ$W, at depth of 600--700~km, see right panel of Fig.~\ref{fig:sc_atlantic}. Such a SC could be a result of subduction of the Pacific plate below the South American plate. The abundances are set to $A_{\text{Th}}=10.5\times10^{-6}\text{ kg/kg}$ and  $A_{\text{U}}=2.7\times10^{-6}\text{ kg/kg}$, with no constraints on total radionuclide mass in the Earth. 
\end{itemize}
For all models, we keep the mantle density identical to the classical case without SC since no indication of density difference in seen in seismic data. The SC matter differs from depleted mantle only by an  increase in heat producing element abundances in the volume specified by a particular SC scenario.

\begin{figure}
\begin{center}
\includegraphics[height=0.25\textwidth]{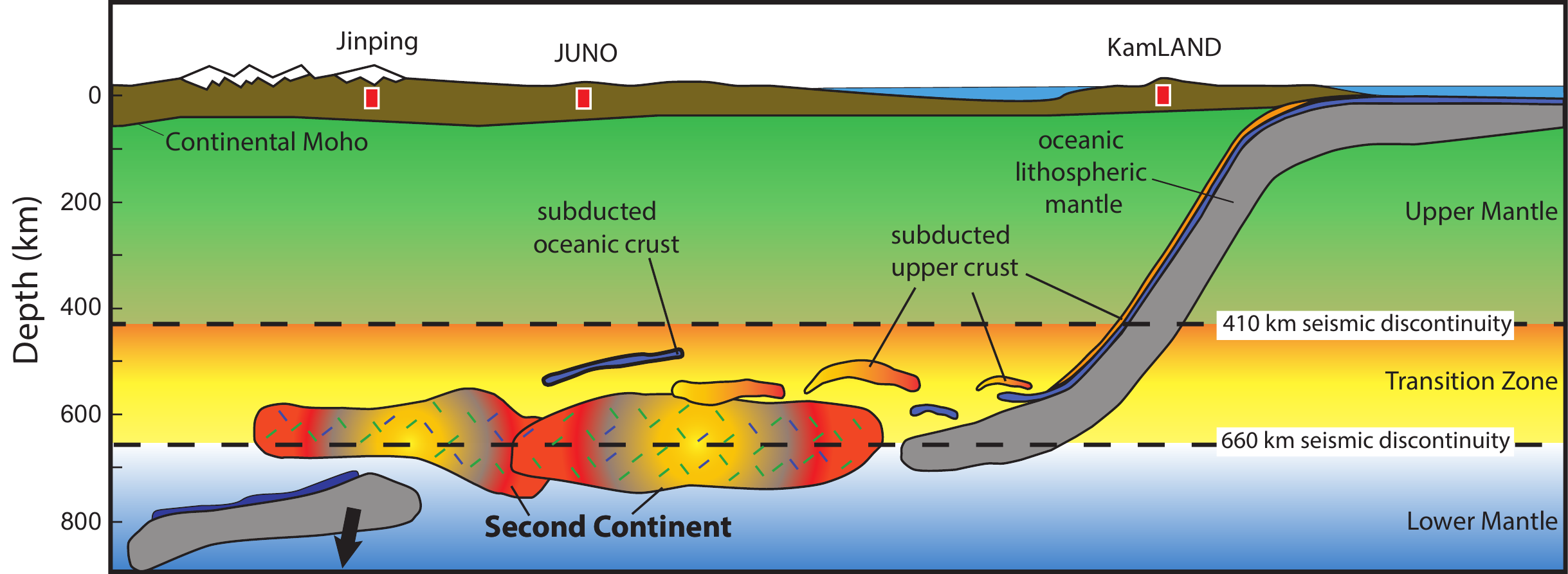}
\includegraphics[height=0.25\textwidth]{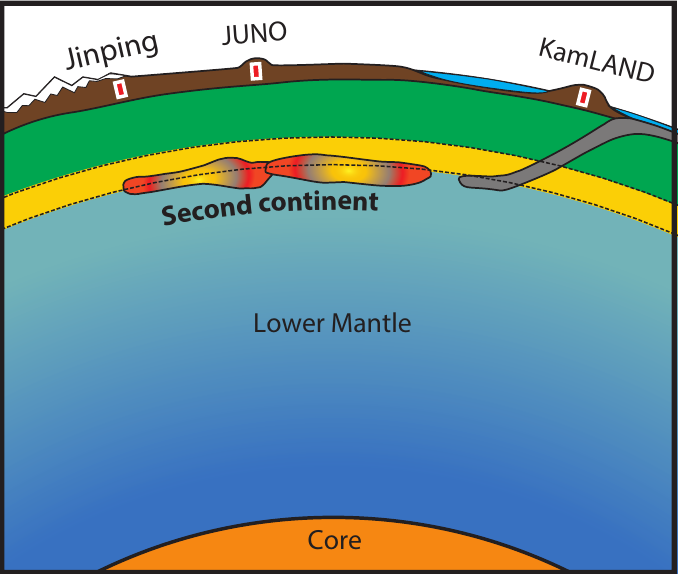}
\caption{\label{fig:cartoon} Left panel shows cartoon illustration of a second continent that formed under eastern China due to Pacific-type subduction occurring along the western margin of the Pacific basin. Picture is an adaptation of Fig.~5. in \citep{KAWAI2013}. The proximity of 3 geoneutrino experiments to this putative second continent are shown as being underground in Japan (KamLAND) and China (JUNO and Jinping). These 3 detectors see approximately the same mantle volume and similar global crust, see right panel, which results in a major reduction in the predicted uncertainties. Local differences in the geoenutrino fluxes for each of these detectors are due to geological differences in the nearby lithosphere (closest 500 km) surrounding the detector; this region contributes approximately half of the detected signal.}
\end{center}
\end{figure}

Model~Ia and Model~Ib represent the old dispersed SC layer while Model~II and Mo\-del~III are examples of potential young SCs. Model~II could be tested by land-based experiments nearby, KamLAND and the upcoming detectors JUNO and Jinping. While JUNO and Jinping will receive a significant flux of geoneutrinos from the SC, KamLAND will be affected only slightly. As the three detectors see the same mantle and similar crust due to their proximity (see Fig.~\ref{fig:cartoon}), a comparison of the measured signals will significantly reduce the impact of the uncertainty of the total Earth's flux prediction on the SC discovery potential.

Model~III could be tested by an ocean bottom movable detector. This mobile instrument will provide flux measurements at multiple locations for signal comparison. Potentially, a future land-based detector located in the ANDES underground laboratory \citep{ANDES} will contribute to test Model~III as well.

\section{Results}
The classical prediction without SC, up to date measurements and expected future relative measurement uncertainty for current and upcoming land-based and ocean bottom experiments are listed in Tab.~\ref{tab:class_results} \citep{SRAMEK}.
\begin{table}[ht]
\begin{center}
\caption{\label{tab:class_results}Prediction of the total geoneutrino flux for land-based and ocean bottom experiments, with their location, latest measurement, if available, and predicted relative uncertainty of the measurement. Predicted flux does not include a contribution from a second continent.}
\begin{tabular}{ccccc}\thickline
Experiment	&	Location	&	\begin{tabular}{c}Predicted \\ flux [TNU] \end{tabular}	& \begin{tabular}{c}Measured\\ flux [TNU] \end{tabular}	& \begin{tabular}{c}Measurement\\ uncertainty\end{tabular} \\ \thickline
KamLAND	&	$36.4^\circ$N $137.3^\circ$E	&	$34.8^{+4.2}_{-4.0}$		&	$30.7\pm7.5$	&	16\% \\\hline
JUNO	&	$22.1^\circ$N $112.5^\circ$E	&	$38.9^{+4.8}_{-4.5}$		&	-	&	6\% \\\hline
Borexino	&	$42.5^\circ$N $13.6^\circ$E	&	$41.4^{+5.1}_{-4.8}$		&	$43.5^{+12.1}_{-10.7}$	&	15\% \\\hline
ANDES	&	$30.2^\circ$S $69.8^\circ$W &	$41.7^{+4.8}_{-4.7}$		&	-	&	5\% \\\hline
SNO+	&	$46.5^\circ$N $81.2^\circ$W 	&	$44.2^{+5.3}_{-5.1}$		&	-	&	9\% \\\hline
Jinping	&	$28.2^\circ$N $101.7^\circ$E	&	$58.5^{+7.4}_{-7.2}$	&	-	&	4\% \\\hline
OBD I	&	$44.0^\circ$S $47.0^\circ$W &	$15.5^{+2.4}_{-2.6}$		&	-	&	10\% \\\hline
OBD II	&	 $44.0^\circ$S $19.0^\circ$W&	$12.7^{+2.4}_{-2.6}$		&	-	&	10\% \\\thickline
\end{tabular}
\end{center}
\end{table}

The additional contribution of the SC can be calculated using Eq.~\ref{eq:fluxcal} and integrating over the SC volume. The abundances of Th and U are taken to be the differences in abundances for an assumed SC and Depleted Mantle.

The combined model of Depleted and Enriched mantle, with abundances from Tab.~\ref{tab:abun}, predicts a total mantle flux to be about 8.1~TNU \citep{SRAMEK}, which is valid for all experiments around the globe that assume a spherically symmetrical mantle composition. The additional geoneutrino flux for Model~Ia, for an experiment at sea level, is about 93~TNU, which leads to an extremely high geoneutrino flux that is already ruled out by current measurements listed in Tab.~\ref{tab:class_results}. Therefore, it is necessary to constrain the amount of radionuclides in the Earth as we did in Model~Ib. The additional signal from a SC for the constrained Model~Ib is 1.6~TNU over the classical mantle flux and this additional signal is significantly smaller than experimental uncertainties of all existing measurements (cf., Tab.~\ref{tab:class_results}). If such a SC described in Model Ib exists, current and upcoming geoneutrino experiments cannot distinguish it from the classical case. More generally, experiments cannot say anything about the vertical distribution of radionuclides in the mantle for models assuming spherical symmetry. In the case of having the mantle's budget of radionuclides located at the base of the mantle, the opposite scenario to Model~Ib, then the expected mantle geoneutrino flux is reduced by about 75\% of the combined model of Depleted and Enriched mantle \citep[see Fig.\,1c of ][]{SRAMEK2013}. This scenario results in about a 2~TNU decrease, assuming an 8~TNU default mantle flux; again, this too is still below the resolution of current and upcoming experiments. 

\begin{figure}
\begin{center}
\includegraphics[width=0.516\textwidth]{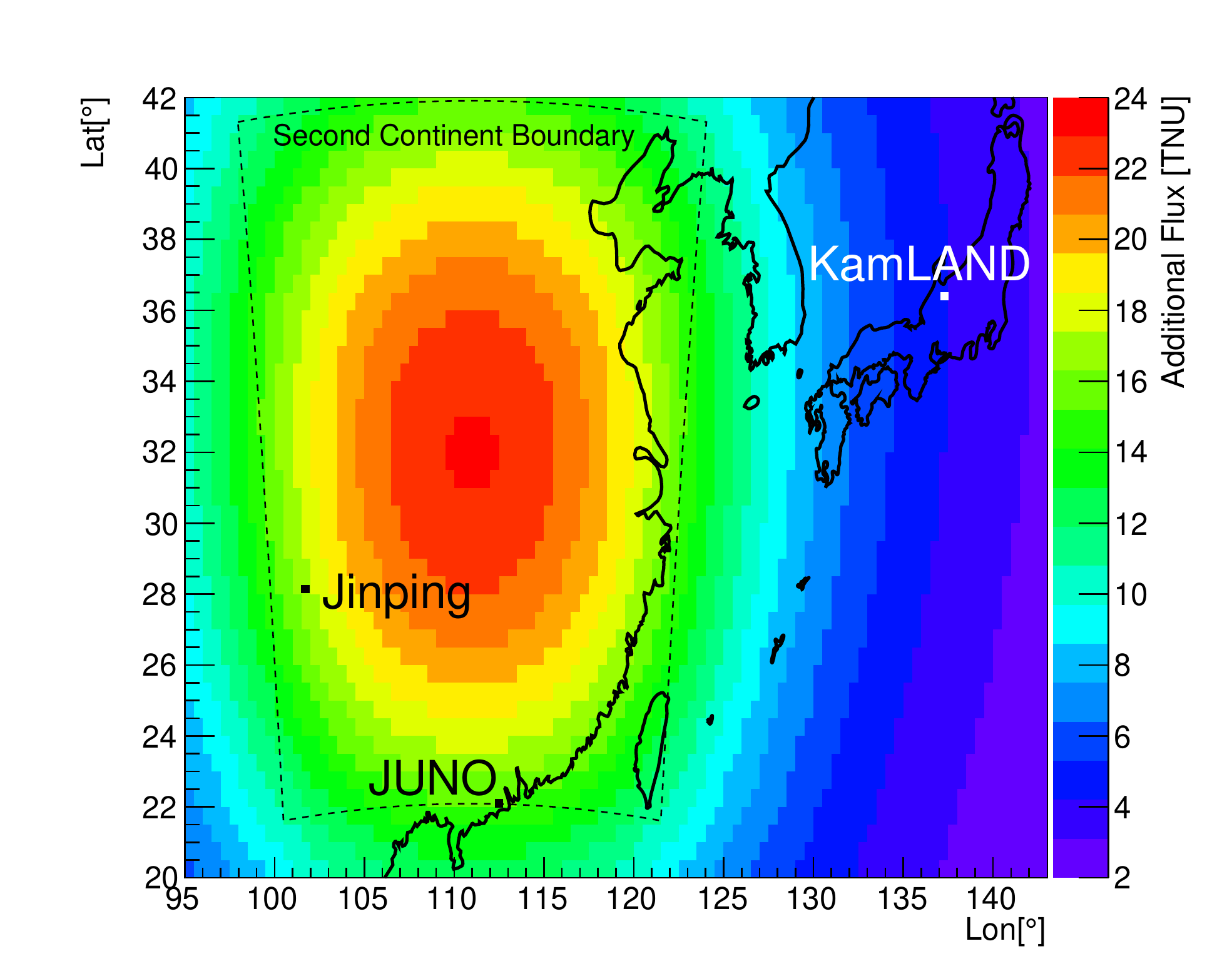}
\includegraphics[width=0.475\textwidth]{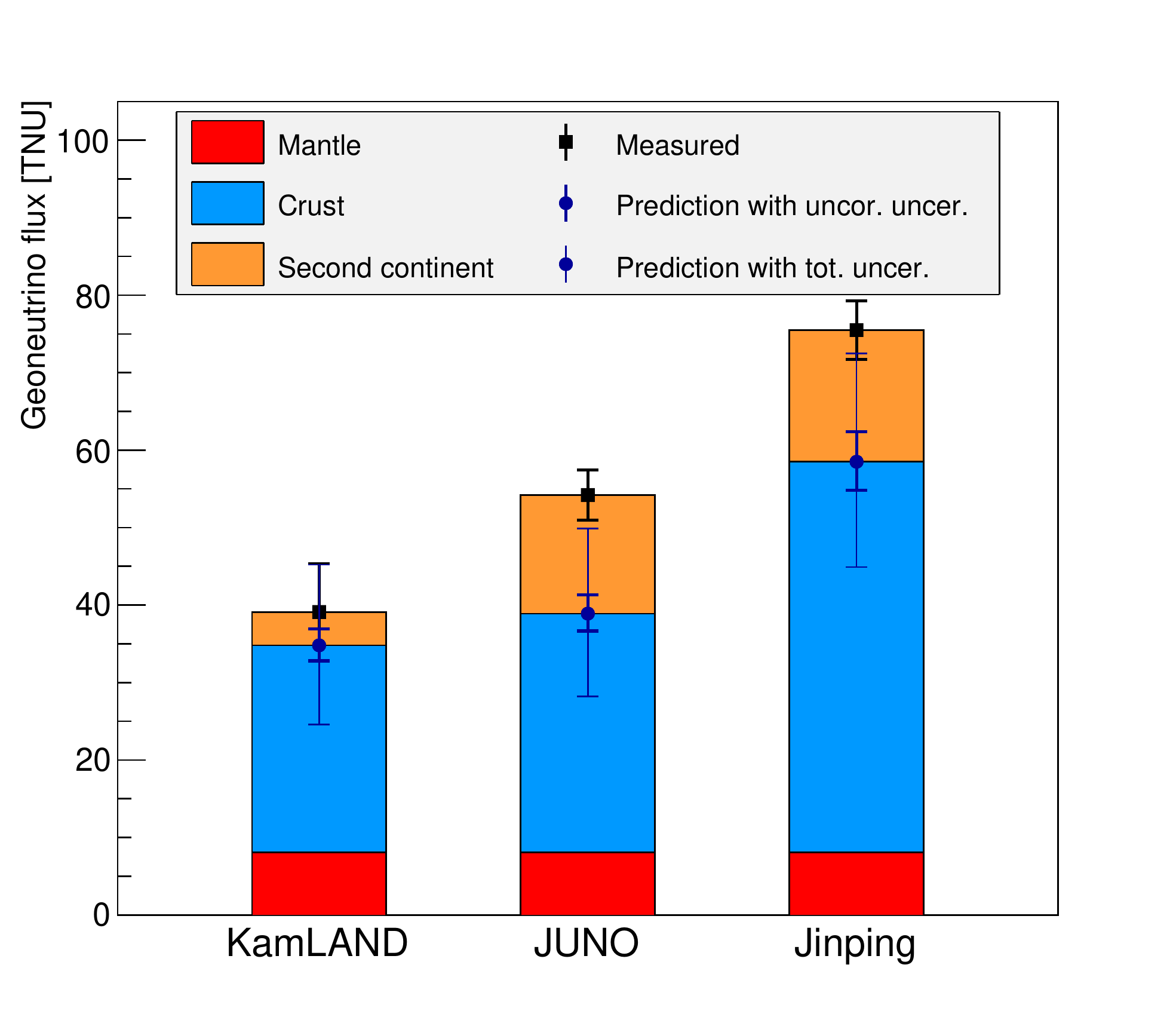}

\caption{\label{fig:sc_asia} Left panel shows the mantle location and surface geoneutrino flux of a SC assumed in Model~II; also identified are the locations of nearby geoneutrino detectors. The geoneutrino flux is reported in TNU, which stands for Terrestrial Neutrino Unit (see text for details). Right panel compares the predicted signal at each experiment broken down into its individual contributions, including the additional geoneutrino signal due to the presence a SC. Since KamLAND will be almost unaffected by the presence of a SC, it provides a benchmark measurement for JUNO and Jinping. These latter two detectors will experience a significant excess in signal over the classical prediction. With the KamLAND measurement as an anchor, the Model~II second continent can be discovered at the $5\sigma$ level.}
\end{center}
\end{figure}

The left panel of Fig.~\ref{fig:sc_asia} shows the additional geoneutrino flux of SC Model~II around its location, highlighting the positions of the nearby geoneutrino detectors. The additional flux at Jinping due to this SC is predicted to be 17.0~TNU, resulting in an overall signal of more than 75~TNU. There would be an additional 15.3~TNU at JUNO due to SC and only 4.3~TNU more at KamLAND. Consequently, there would be a significant increase in the signal at Jinping and JUNO, while KamLAND would receive only a small additional contribution from the SC. These results allows us, in principle, to discover a SC under eastern China. The crucial feature of the modeling here is the comparison of signals between nearby experiments, which cancels out potential uncertainties that come with the geological model parameters (i.e., assumed global crustal and mantle models). The right panel of  Fig.~\ref{fig:sc_asia} shows the relative contributions of the mantle, continental crust, and second continent to the signals at the three detectors, assuming a shared 8.1~TNU mantle signal. We highlight the measured flux with an expected uncertainty from Tab.~\ref{tab:class_results} to predict the classical case without a second continent. If one takes into account the larger uncertainty, which includes the effect of various mantle models, where we assume 1.9~TNU flux from Low-Q mantle and 14.3~TNU flux from High-Q mantle, we cannot test the presence of a SC. However, since these 3 detectors are in the same part of the world (with a maximum surface separation distance comparable to a mantle depth), they will see the same mantle and thus the uncertainty of the mantle model can be neglected. In addition, the prediction for the crust will be highly correlated and thus only part of its uncertainty will matter. 

In order to quantify the discovery potential, we calculate a $\chi^2$ value defined as:
\begin{equation}
\label{eq:chi2}
\chi^2=\sum_{i=1,2,3}\left(  \frac{S_{\text{wSC}}^i-S_{\text{woSC}}^i}{\sqrt{(\sigma_m^i)^{2}+(\sigma_p^i)^{2}}} \right)^2=\sum_{i=1,2,3}\left(  \frac{S_{\text{SC}}^i}{\sqrt{(\sigma_m^i)^{2}+(\sigma_p^i)^{2}}} \right)^2,
\end{equation}
where $S_{\text{wSC}}^i$ is total signal, including contribution from SC, $S_{\text{woSC}}^i$ is total signal without SC contribution, and $S_{\text{SC}}^i$ is the additional signal from a SC for i-th experiment. The measurement uncertainty $\sigma_{m}^i$ is the product of the total signal of the SC and the expected future relative uncertainty from Tab.~\ref{tab:class_results}. The impact of the prediction uncertainty will be significantly reduced due to the high correlation between experimental sites. We assume the uncorrelated part of the prediction uncertainties among experiments $\sigma_p^i$ to be 50\% of the ones listed in Tab.~\ref{tab:class_results}. We assume that $\chi^2$ defined in Eq.~\ref{eq:chi2} follows $\chi^2$ distribution for one degree of freedom. In that case, we can express the confidence level of SC discovery by the number of standard deviations as $\sqrt{\chi^2}$. We have found that Model~II SC can be discovered at the $5.0\sigma$ level. The potential drops to $2.7\sigma$ with half of the content of U and Th in the SC. We note that these discovery potentials are directly linked to the location, size, and Th and U abundances of the SC set by Model~II. Nevertheless, qualitative conclusions are valid with similar SC: there is a big potential of discovery of the SC under China by these upcoming geoneutrino experiments.

\begin{figure}
\begin{center}
\includegraphics[width=0.516\textwidth]{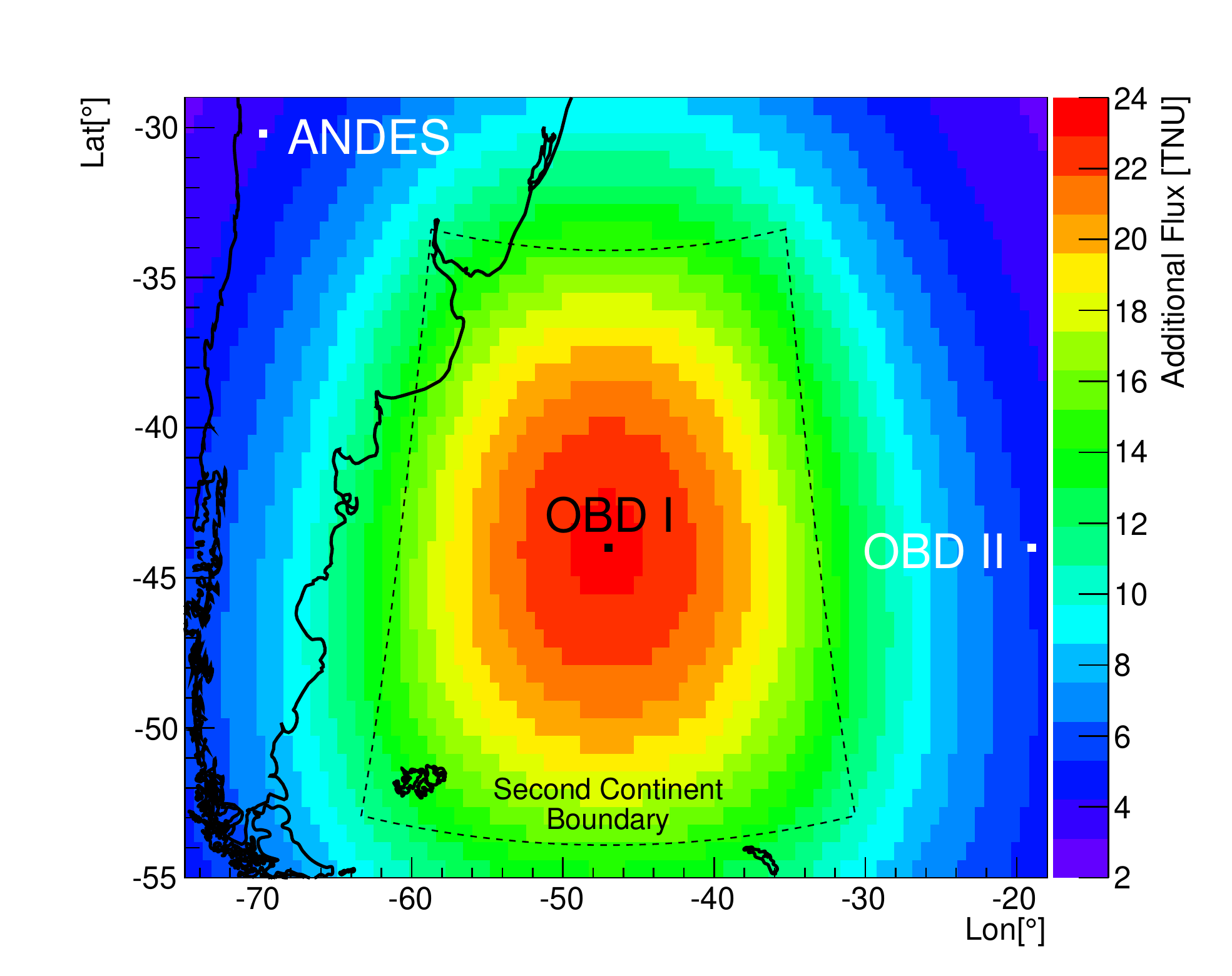}
\includegraphics[width=0.475\textwidth]{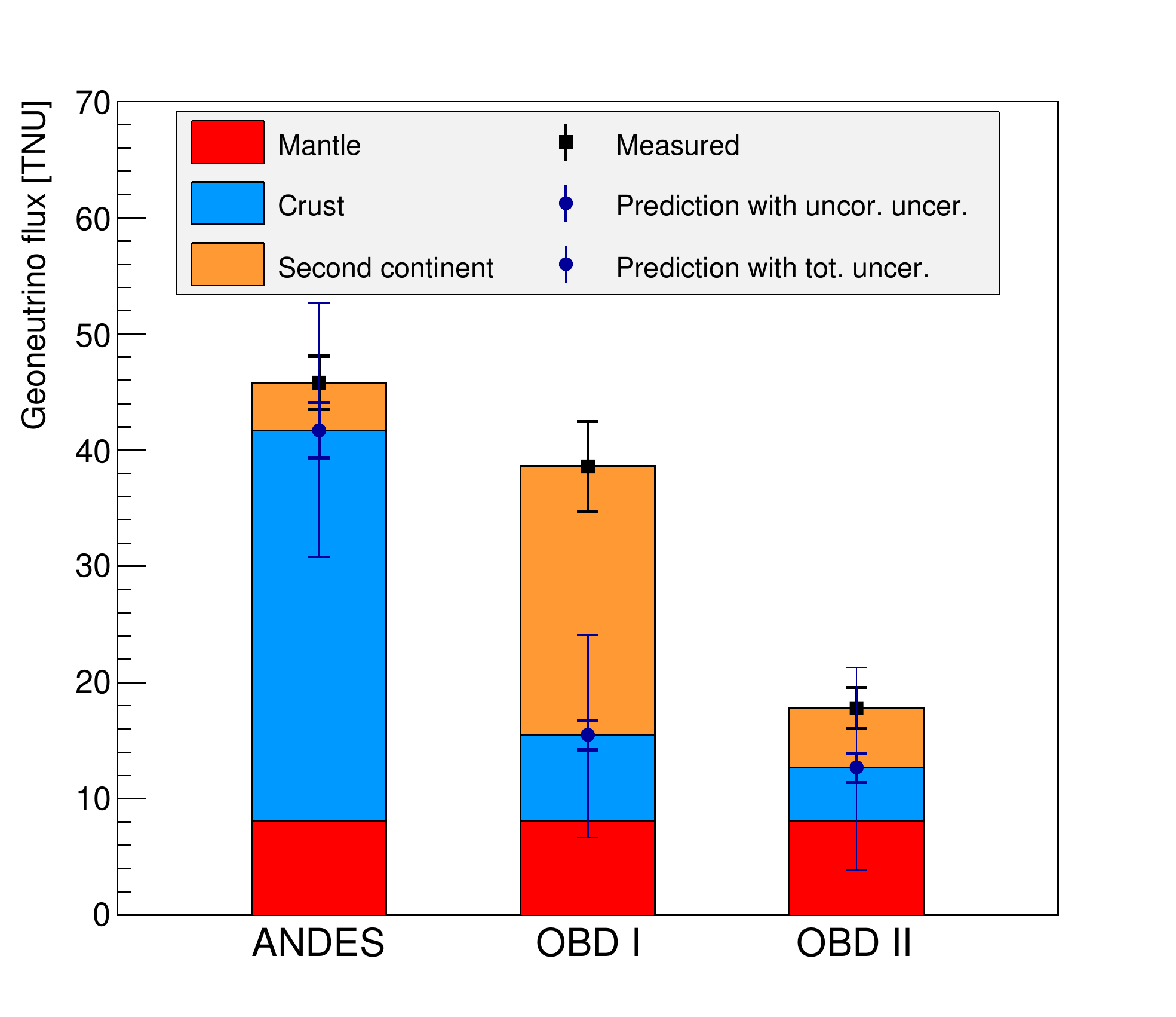}
\caption{\label{fig:sc_atlantic} Left panel shows the mantle location and surface geoneutrino flux of a SC assumed in Model~III; also identified are the locations of nearby geoneutrino detectors, specifically the proposed OBD (Ocean Bottom Detector) locations and the future site of ANDES laboratory. The geoneutrino flux is reported in TNU, which stands for Terrestrial Neutrino Unit (see text for details). Right panel compares the predicted signal at each experiment broken down into its individual contributions, including the additional geoneutrino signal due to the presence a SC. OBD~I is position above the center of second continent, while the position of OBD~II was selected to be marginal to the SC. The proposed ANDES detector has the potential to discover such a second continent as modeled here. Collectively, these 3 detectors can be used to discover a Model~III second continent at the $6.2\sigma$  level.}
\end{center}
\end{figure}

The additional geoneutrino flux of a SC for Model~III is shown on the left panel of Fig.~\ref{fig:sc_atlantic}. Two locations of an OBD (Ocean Bottom Detector) can be compared in order to reduce the effect of the prediction uncertainty. OBD~I is centrally located above SC experiencing an additional signal of 23.1~TNU. OBD~II is located 2000~km off the SC center to suppress the SC contribution while being reasonably nearby to see the same mantle and crust. The measurement of an Atlantic SC can be potentially detected by a land-based detector located at the ANDES underground laboratory. Together, the discovery potential is $6.2\sigma$. We observe that a single OBD~I measurement can alone provide the evidence of a SC despite the large prediction uncertainty due to the ideal location of OBD~I above the center of a hypothesized SC.

\section{Conclusions}
The emerging field of particle geophysics (including studies in geoneutrinos, neutrino oscillations, neutrino absorption, and muography) offers its existing technologies to interrogate independently a range of problems in geoscience. Using a suite of detectors that are co-located on the scale of a mantle depth, we propose using differences in the measured flux of geoneutrinos at these detectors to test for the existence of a second continent \citep{KAWAI2013, SAF_MARU} located in the mantle Transition Zone beneath eastern China. This second continent, with its high content of radionuclides, would be a bright geoneutrino emitter and readily detectable with existing technologies.
 
Here we evaluated several simple, second continent models, which have been proposed. Based on existing technologies, we cannot distinguish the model of an ancient (i.e., $>$1~Ga) second continent, which has evolved to a uniform layer globally encircling the base of the Transition Zone. In general, geoneutrino measurement cannot reveal spherically symmetric geoneutrino sources in the mantle, unless the uncertainty of the measurement, as well as prediction, improves significantly.
 
In contrast, we can successfully identify geoneutrino bright sources coming from young (i.e., $<$1~Ga) second continents that are formed by the entrainment of continental upper crust in subduction zones and aggregated into compact domains at the base of the Transition Zone.  Published tectonic models envisage that ocean-continent convergence provide a steady supply of upper continental crust to landward-projected, accumulation zones that are density stabilized at the base of the Transition Zone and in the uppermost lower mantle. These accumulation zones contain a considerable amount of heat producing elements that are proposed as drivers (i.e., providing thermal energy and volatiles) of present-day magmatism found above second continents. Our evaluation considered two different Australia-size young second continents: one located beneath eastern China and the second beneath the south Atlantic, eastward of South America. Following literature recommendations \citep{KAWAI2013, SAF_MARU}, these second continents, with their crustal abundances of radioactive heat producing elements, provide enhanced geoneutino fluxes that allow for their discovery. A second continent beneath eastern China will readily be detectable with the land-based detectors KamLAND, Jinping, and JUNO. A second continent beneath the south Atlantic will readily be detectable by a proposed ocean-based, movable geoneutrino detector along with the land-based ANDES detector.

We also highlight that a Chinese second continent should influence the regional heat flux in China. The regionally averaged surface heat flux for eastern China (60~mW/m$^2$) \citep{GAO19981959} is slightly lower, but comparable to the global average, continental heat flux of 65~mW/m$^2$ \citep{POLLACK1993,jaupart:2015tg,DAVIES2013} and 71~mW/m$^2$ \citep{DAVIESDAVIES2010}. Thus, given the existence of a Chinese second continent, then its existence will lead to an enhanced Moho heat flux, and necessarily a reduced crustal heat production. This latter prediction can be independently tested with combined geological and geoneutrino studies.

\acknowledgments
BR is grateful for support of Comisi\'{o}n Nacional de Investigaci\'{o}n Cient\'{i}fica y Tecnol\'{o}gica [POSTDOCTORADO-3170149]. O\v{S} gratefully acknowledges Czech Science Foundation support [GA\v{C}R 17-01464S] for this research. WFM gratefully acknowledges National Science Foundation support [NSF-EAR1650365] for this research.

\bibliography{sc_references}

\end{document}